\documentstyle[12pt]{article}
\begin{document}
\title{THE GIBBS PARADOX, THE LANDAUER PRINCIPLE AND THE IRREVERSIBILITY ASSOCIATED WITH TILTED OBSERVERS}
\author{L. Herrera$^1$\thanks{On leave from UCV, Caracas, Venezuela, e-mail: lherrera@usal.es}\\
{$^1$Instituto Universitario de F\'isica
Fundamental y Matem\'aticas},\\ {Universidad de Salamanca, Salamanca 37007, Spain}}
\maketitle

\begin{abstract}
 It is well known that, in the context of General Relativity, some spacetimes,  when described by a congruence of comoving observers,  may consist in a distribution of a perfect (non--dissipative) fluid, whereas the same spacetime as seen by a ``tilted'' (Lorentz--boosted) congruence of observers, may exhibit the presence of dissipative processes. As we shall see, the appearence of entropy producing processes  are related to the tight dependence of entropy on the specific congruence of  observers. This fact is well illustrated by the Gibbs paradox. The appearance of such dissipative processes, as required by the Landauer principle, are necessary, in order to erase the different amount of information stored by comoving observers, with respect to tilted ones.
\end{abstract}

\section{INTRODUCTION}
{\it ``Irreversibility is a consequence of the explicit introduction of ignorance into the fundamental laws.'' } M. Born
\medskip

Observers play an essential role in any physical theory. This is  is particularly true in Thermodynamics and in General Relativity.

Indeed, in this latter theory, it is well known  that  a variety of line elements may satisfy the Einstein equations for different (physically meaningful) stress--energy tensors (see \cite{1}--\cite{12T} and references therein). This ambiguity in the description of the source may  be related, in some cases,  to the arbritariness in the choice of the four--velocity in terms of which the energy--momentum tensor is split.

The above mentioned arbitrariness, in its turn, is related to the well known fact, that different congruences  of observers would assign different four--velocities to a given fluid distribution. We have in mind here, the situation when one  of  the conguences corresponds to comoving observers, whereas the other is obtained by applying a Lorentz boost to the comoving observers. 

For example, in  the case of the zero curvature FRW model, we have  a perfect fluid solution for observers at rest with respect to the timelike congruence defined by the eigenvectors of the Ricci tensor, whereas  for observers moving relative to the previously mentioned congruence of observers, it can also be interpreted as the exact solution for a viscous dissipative fluid  \cite{4}. It is worth noticing that the relative (``tilting'') velocity between the two congruences may be related to a physical phenomenon such as the observed motion of our galaxy relative to the microwave background radiation \cite{9}.

Thus,   zero curvature FRW models as described by ``tilted'' observers, will exhibit a dissipative fluid and  energy--density inhomogeneity, as well as different values for  the expansion scalar and the shear tensor, among other differences, with respect to the ``standard''  (comoving) observers (see \cite{4} for a comprehensive discussion on this example). 

The same phenomenon appears in the tilted versions of the Lemaitre--Tolman--Bondi  (LTB) \cite{25}--\cite{27} (see \cite{38}),  the Szekeres spacetimes \cite{1s, 2s} (see \cite{2S}), and in many other circumstances (see \cite{t1}--\cite{ t5}  and references therein) .

At this point, we should mention that in the past it has been argued that  dissipative fluids (understood as fluids whose energy--momentum tensors present a non--vanishing heat flux contribution), are not necessarily incompatible with reversible processes (e.g see \cite{14T}--\cite{16T}). 

In the context of the standard Eckart theory \cite{17T}, a necessary condition for the compatibility of an imperfect fluid with vanishing entropy production (in the absence of bulk viscosity) is  the existence of a conformal Killing vector field CKV) $\chi^\alpha$ such that $\chi^\alpha =\frac{V^\alpha}{T}$ where $V^\alpha$ is the four--velocity of the fluid and $T$ denotes the temperature. In the context of   causal  dissipative  theories, e.g. \cite{18T}--\cite{23T}, the existence of such CKV is also necessary for an imperfect fluid to be compatible with vanishing entropy production (see \cite{38}). 

However, a much more carefull analysis of the problem readily shows, that  the compatibility of reversible processes and  the existence of dissipative fluxes becomes trivial if a constitutive transport equation is adopted, since in this latter case such compatibility  forces the heat flux vector to vanish as well. In other words, even if {\it ab initio} the fluid is assumed imperfect (non--vanishing heat flow vector) the imposition of the CKV and the vanishing entropy production condition may cancel the heat flux,  once a  transport equation is assumed (see \cite{NDT} for a detailed discussion on this point).

In other words, in the presence of a CKV of the kind mentioned before, the  assumption of a  transport equation whether in the context of the Eckart--Landau theory, or a causal theory, implies that  a vanishing entropy production leads to a vanishing heat flux vector. Therefore, under the conditions above,   the system is not only reversible but also non dissipative.

Furthermore, since  neither LTB nor the Szekeres spacetimes admit a CKV, we may safely conclude that the heat flux vector  appearing in these cases, is associated to truly (entropy producing) dissipative processes.

To explain the origin of such processes, is the main purpose of this work.

\section{COMOVING AND TILTED OBSERVERS}
Let us consider  a  congruence of observers which  are  comoving with a dissipationless dust distribution, then the four--velocity for that congruence, in some globally defined coordinate sytem,   reads
\begin{equation}
v^\mu =(1,0,0,0).
\label{vMc}
\end{equation}

In order to obtain the four--velocity corresponding to the tilted congruence (in the same globally defined coordinate system), one proceeds as follows.

We have first to perform a (locally defined) coordinate transformation to the  Locally Minkowskian Frame (LMF). Denoting  by $L _\mu ^\nu$  the local coordinate transformation matrix, and  by $\bar v^\alpha$ the components of the four velocity in such LMF, we have:
\begin{equation}
\bar v^\mu = L^\mu_\nu v^\nu.
\label{V}
\end{equation}

Next, let us perform  a Lorentz boost from the LMF associated to  $\bar v^\alpha$, to the (tilted)  LMF with respect to which a fluid element is moving with some, non--vanishing, three--velocity.

Then the four--velocity in the tilted LMF is defined by:
\begin{equation}
\tilde v_{\beta}=\Lambda^\alpha_\beta \bar v_\alpha,
\label{1}
\end{equation}
where  $\Lambda^\alpha_\beta$ denotes the Lorentz matrix.

Finally, we have to perform a  transformation from the tilted LMF, back  to the (global) frame associated to  the line element under consideration. Such a transformation, which obviously only exists locally,  is defined by the inverse of  $L _\mu ^\nu$, and produces the four--velocity of the tilted congruence, in our globally defined coordinate system, say $V^\alpha$.

Let us now consider a given spacetime, which according to comoving observers, is sourced by a dissipationless dust distribution, so that the energy momentum--tensor reads
\begin{equation}
T^{C}_{\mu\nu}= \mu_{C} v_\mu v_\nu ,
\label{Tpo}
\end{equation}
where $C$ stands for comoving and  $\mu_{C}$ denotes the energy density, as measured by the comoving observers.

However for the tilted congruence we may write
\begin{equation}
{T}^{T}_{\alpha\beta}= (\mu_{T}+P) V_\alpha V_\beta+P g _{\alpha \beta} +\Pi_{\alpha \beta}+q_\alpha V_\beta+q_\beta V_\alpha,
\label{6bis}
\end{equation}

where $T$ stands for tilted, and $\mu_{T}$, $q_\alpha$   $P$  and $\Pi_{\alpha \beta}$ denote the energy density, the heat flux, the isotropic pressure, and  the anisotropic tensor, as measured by the tilted observers. 

Obviously, both  energy--momentum tensors are exactly  the same, since the metric is the same and therefore the Einstein tensor is the same,  however the way  in which the energy--momentum tensor is split, is not the same. This simple fact opens the possibility (for tilted observers) to obtain an energy--momentum tensor which describes a quite different picture from the one obtained by comoving observers. In the same order of ideas, it should be emphasized that the kinematical variables (four--acceleration, expansion scalar, shear tensor, vorticity tensor), being defined in terms of the four--velocity, will also differ from their values as measured by the comoving observers.

Among the differences appearing in the tilted congruence, with respect to the comoving one, there is one which rises  the most intriguing question, namely: how it is possible  that tilted observers may detect irreversible processes, whereas comoving observers describe an isentropic situation ?

As we shall see, the answer  to the above question is closely related to the  fact that the definition of entropy  is highly observer dependent, as illustrated, for example, by the Gibbs paradox.

\section{THE GIBBS PARADOX, THE LANDAUER PRINCIPLE  AND THE DEFINITION OF ENTROPY}

Entropy  is a measure of how much is not known (uncertainty). Also known, although usually overlooked,  is the fact  that physical objects do not have  an intrinsic uncertainty (entropy) (see \cite{adami} for an enlightening discussion on this issue). 

The ``subjective'' nature of the concept of entropy is brightly illustrated by the Gibbs paradox. In its simplest  form, the paradox appears from the consideration of a box divided by a wall in two identical parts, each of which is filled with an ideal gas (at the same pressure and temperature). Then if  the partition wall is removed, the gases of both parts of the box will mix. 

Now, if the gases from both sides are distinguishable, the entropy of the system will rise, whereas if they are identical there is no increase in entropy. This leads to the striking conclusion that irreversibility (and thereby entropy), depends on the ability of the observer to distinguish, or not, the gases from both sides of the box. In other words, irreversibility would depend on our knowledge of physics \cite{bais}, confirming thereby our previous statement  that physical objects are deprived of intrinsic entropy. It can only be defined {\bf after} the number of states that can be resolved by  the measurements, are established.  The anthropomorphic nature of entropy has been brought out and discussed in detail by Jaynes \cite{J}.

Let us now turn back to our  comoving and tilted observers.

 If a given physical system is studied by a congruence of comoving observers, this implies at once that the three--velocity of any given fluid element is automatically assumed to vanish, whereas for the tilted observers this variable represents an additional degree of freedom. In other words, the number of possible states in the latter case is much larger than in the former one. 

Since  for the comoving observers  the system is dissipationless, it is clear that the increasing of entropy, when passing to the tilted congruence, should imply the presence of dissipative (entropy producing) fluxes, in the  tilted congruence. 

It is instructive to take a look on this issue from a different perspective, by considering the transition from the tilted congruence to the comoving one.
According to the Landauer principle, \cite{lan} (also referred to as the Brillouin principle \cite{3B}--\cite{8B}), the erasure of one bit of information stored in a system requires the dissipation into the environment of a minimal
amount of energy, whose lower bound is given by 
\begin{equation}
\bigtriangleup E=kT \ln2,
\label{lan1}
\end{equation}
where $k$ is the Boltzmann constant and $T$ denotes the  temperature of the environment. 

In the above, erasure, is just a reset operation restoring the system to a specific state, and is achieved by means of an external agent. In other words, one can decrease the entropy of the system by doing work on it, but then one has to inccrease the entropy of another system (or the environment). 

Thus, Landauer principle is an expression of the fact that logical irreversibility necessarily implies thermodynamical irreversibility.

Now, when passing from the tilted to the comoving congruence, a decrease of entropy occurs, but we have not any external agent, and therefore such a decrease of entropy  is accounted by the dissipative flux observed in the tilted congruence (we recall  that in the comoving congruence the system is dissipationless).

 The point is, that passing from one of the  congruences  to the other, we usually overlook the fact that  both congruences of observers store different amounts of information. Here resides the clue to resolve the quandary mentioned above, about the presence or not of dissipative processes, depending on the congruence of observers, that carry out the analysis of the system.

Before concluding this section, three remarks are in order:
\begin{itemize}
\item The main issue discussed in this work, namely: the presence or not of dissipative processes, depending on the congruence of observers, that carry out the analysis of the system, will remain for any theory of gravity. However, specific details of the dissipative processes observed by the tilted observers, will  depend on the  theory of gravity under consideration.
\item The discussion about the entropy budget of the universe, is of the utmost relevance (see \cite{nc} and references therein),  because its increase is associated with all possible irreversible processes, on all scales. However, in that reference, as well as in the references therein, the issue under consideration is the estimate of entropy as observed by  {\it one given} congruence of observers. The main point of our work is to stress how  (and why)  any of these estimates, changes when it is evaluated by  {\it different} congruences of observers. 
\item  It goes witouh saying that, in the context of a covariant theory of gravity  (such  as GR), a covariant definitions of entropy should be invoked. Such a definition can be found in the context of different relativistic dissipative theories (see for example \cite{18T}--\cite{23T}). However we have not made use of them in the text, which explain why we did not  refer to this particular issue.
\end{itemize}

\section{Conclusions}
We may summarize the main issues addressed in this letter, in the following  points:
\begin{itemize}
\item  Uncertainty (entropy) is tightly dependent on the observer.
\item Comoving and tilted observers, store different amounts of information.
\item According to the Landauer principle, erasure of information is always accompanied by dissipation (there is a price to forgetting).
\item The detection of dissipative processes by tilted observers, in physical systems which  are  described by  comoving observers, as    perfect fluids, becomes intelligible at the light of the three previous comments.
\end{itemize}
\section{Acknowledgments}
This  work  was partially supported by the Spanish  Ministerio de Ciencia e
Innovaci\'on under Research Projects No.  FIS2015-65140-P (MINECO/FEDER).

\end{document}